# Palladium: A localised paramagnetism

A. Hernando*,[1,2], B. Sampedro[1,2], M. A. García[1,2], P. Marín[1,2], G. Rivero[1,2], P. Crespo[1,2]

and J. M. González[1,3]

[1] *Instituto de Magnetismo Aplicado. RENFE-CSIC-UCM. P.O. Box 155. Las Rozas, Madrid 28230, Spain*

[2] *Dpto. de Física de Materiales. UCM, Madrid, Spain*

[3] *Instituto de Ciencia de Materiales. CSIC. Canto Blanco, Madrid, Spain*

* Corresponding author: ahernando@renfe.es

**Abstract**

We report on the origin of ferromagnetic like behaviour observed for 2.4 nm size Pd nanoparticles. Localised magnetic moments on metallic surfaces have been recently shown to induce orbital motion of the itinerant electrons via spin-orbit coupling. As a result of this coupling the magnetic anisotropy is enhanced and the surface magnetic moments can be blocked up to above room temperature. Since Pd has been customary treated as a paradigmatic itinerant system, localisation of magnetic moments at its surfaces was not initially expected. However, it is shown, through the experimental thermal dependence of both magnetic susceptibility and Hall resistance, that magnetism of bulk Pd, is a localised paramagnetism and, consequently, can give rise to permanent magnetism at its surface. Such surface permanent magnetism is experimentally observed only when the percentage of surface moments is outstanding as is the case for nanoparticles.

**Text**

It has been recently observed and reported that metallic nanoparticles (NPs) of 1-2 nm size can exhibit permanent magnetism at room temperature. These observations were performed for Pd [1-3] and also later for thiol capped Au [4-5] NPs. Whereas Pd presents ferromagnetic instability, Au is a diamagnetic metal. However, the thermal dependence of magnetization was found to be the more intriguing characteristic of this new type magnetic behaviour. In both cases, the magnetic moment at constant applied field does not change with temperature between 5 and 300 K. The presence of hysteresis at room temperature in so small NPs implies that the magnetic moments of NPs are blocked by a giant anisotropy. Gambardela et al [6], have observed, by using circular magnetic dichroism, that even single atoms of Co deposited onto Pt (111) surface exhibit giant anisotropy, of the order of $1\,meV/atom$.

For the case of Au NPs, the binding with a sulphur atom induces and localises a surface magnetic moment in the d band of Au for which spin-orbit coupling is of the order of 0.4 eV. The giant orbital magnetic moments observed in thiol capped Au films [7] have been explained also in terms of the spin-orbit coupling that promote the induction of high orbital momentum of quasi-free electrons localised around the thiol binding [8]. The orbital momentum induced in the conduction electrons gives rise to a strong enhancement of the perpendicular anisotropy of the localised moment. Note that in the experiments carried out for Co adatoms [6] and for thiol capped gold films and NPs the magnetic moments were well localised. This seems not to be the case for Pd whose magnetic properties were analysed during the last sixty years in the framework of itinerant magnetism.

Free atoms of Pd exhibit noble gas configuration. However, the overlapping between *4d* and *5s* band leads to a *$4d^{10-\xi} 5s^{\xi}$* occupation for those Pd atoms embedded in its characteristic fcc lattice. Paramagnetism can be only originated from incomplete bands i. e. from holes at the *4d* band with number per atom, *$\xi(T)$*, and from electrons at the *5s* band with the same number. Its high density of states at the Fermi level, *$1,23\ states \cdot eV^{-1} \cdot spin^{-1} \cdot atom^{-1}$*, has been reported to be the cause of an enhanced paramagnetism (exchange Stoner enhancement factor of about 10). This enhancement factor was necessarily introduced to account for the extremely high experimental susceptibility, *$\chi = 10^{-3}$*, difficult to explain in a pure Pauli paramagnetism framework [9]. The susceptibility of Pd presents, besides its large value, a peculiar thermal dependence exhibiting a maximum close to 80K [9,10]. Furthermore, the susceptibility follows the Curie-Weiss (C-W) law, for temperatures above 150 K. This thermal dependence, anomalous for a system obeying the Fermi-Dirac statistics, was tried to be explained by using the more sophisticated arguments but always within the framework of the magnetism of itinerant electrons [11-18]. For instance, it was claimed that a singularity in the density of states near the Fermi surface combined with exchange enhancement could provide a possible explanation of the C-W behaviour for Pauli paramagnetism [19]. It is to note that the susceptibility of Ni just above its Curie temperature, 630 K, and that of Pt above 300 K, also obey a C-W law so that such thermal dependence does not seem to be a consequence of any particular and subtle characteristic of Pd band topology but a more general property of some transition metals.

Localisation as cause of permanent magnetism and giant anisotropy development in NPs seems difficult to be claimed for the case of Pd, metal for which customary itinerant magnetism was considered. In order to inquire closer in the origin of the Pd NPs behaviour we have first performed an experimental analysis of bulk Pd magnetism. This study led to conclude that, despite the general acceptance, the main contribution to bulk Pd paramagnetism comes from localised magnetic moments. Secondly, the presence of localised magnetic moments at the surfaces combined with the enhanced spin-orbit coupling is shown to account for the permanent magnetism observed in Pd NPs. It will be shown that the thermal increase of $\xi(T)$ combined with the localised character of the $d$ band holes and the itinerant character of the $s$ electrons, determine the Pd magnetic behaviour.

Magnetization of bulk polycrystalline Pd (Aldrich product number 20393-9, lot number 02616DC, with impurities content below *0,001 at.%*) has been thoroughly measured for a broad range of field and temperature, by using SQUID magnetometry, as shown in Figure 1. At temperatures above 150 K, the paramagnetic susceptibility obeys the Curie-Weiss law illustrated in Figure 2 and corresponding to a magnetic moment, $m$, of 1.5 $\mu_B$ per atom.

Hereinafter we will assume that C-W behaviour is a footprint of localised magnetism. Therefore, the experimental results indicate that the paramagnetic behaviour observed at high temperatures corresponds mainly to an important contribution of localised electrons. It is understood that an electron can be considered magnetically localised when it belongs to a band for which the hopping is lower that the intra-atomic exchange. This condition for Pd can only hold for the holes, $\xi(T)$, of some *d*-sub-band, narrower than the broad *s*-band. Therefore, the expected magnetic susceptibility should contain besides the experimental C-W contribution, a Pauli contribution of itinerant *s*- electrons with magnetic moment per atom $\mu_B \cdot \xi(T)$. Note that the localised atomic magnetic moment is also $m(T) = \mu_B \cdot \xi(T)$, provided the quenching of angular momentum.

Remind that the Curie susceptibility, $\chi$, is given by [20].

$$\chi = \frac{N \cdot m^2(T)}{3 \cdot k_B \cdot (T - \Theta)} \qquad (1)$$

where $N$ is the number of atoms per unit volume and $k_B$ the Boltzman constant. The itinerant contribution due to delocalised *s*-band electrons obeys a relationship as that given by (1), with the same magnetic moment per atom, $\mu_B \cdot \xi(T)$, but in which $T_F$, of the order of $10^5$ $K$, appears instead of $T$. This contribution can be neglected in a first approximation, as experimentally confirmed. It is important to remark that the localised magnetic moment per atom can be fractional due to the overlapping between *d* and *s* bands.

Deviation from the C-W law observed at temperatures below 150 K could be understood as a consequence of the increase of $m(T)$ as temperature rises from 5 to 150 K, temperature at which $m(T)$ reaches its saturation value. Such increase of the number of localised holes, $\xi(T)$, at *d* sub-bands, must be induced by electron thermal transitions towards more delocalised *s*-band states. In other words Pd would thermally behave as an intrinsic semiconductor with zero gap width, due to the *3d-4s* band overlapping. According to this statement, as $T$ rises, the progressive enhancement of $m(T)$ may be counterbalanced by the increasing thermal disorder so giving rise to the susceptibility maximum.

$m(T)$, as obtained from the difference between the experimental susceptibility and the susceptibility corresponding to the high temperature C-W fitting has been plotted in Figure 3. Consider that $\theta$ can only be obtained from the fitting of the C-W law (1) at temperature higher than that at which $m(T)$ reaches its constant saturation value. Therefore, the $m(T)$ dependence plotted in Figure 3 can be considered as estimation since it was obtained by considering a constant $\theta$ for all the measuring temperature range.

The assumed increase of $m(T)$, can be tested by considering that it must be associated with an increase of the itinerant *s*-band electrons, $\xi(T)$, therefore, giving rise to a decrease of the Hall resistance. When the mobility of the $\xi(T)$ localised heavy holes is neglected - in comparison to the high mobility of the $\xi(T)$ light electrons- the Hall resistance can be directly related to the $\xi(T)$, [21]. The drop of the Hall resistance with increasing $T$ strongly suggests an increase of the number of carriers. We have also drawn, in Figure 3, the number of carriers $\xi(T)$, as a function of temperature as obtained from the Hall resistance results reported in [22]. It is observed that both magnetic moment and number of carriers per atom follow the same thermal dependence. Moreover, for $T = 150K$, $\xi(T)$ is 1.4 electrons per atom while the magnetic moment value is very similar: 1.5 $\mu_B$ per atom. The quantitative and qualitative agreement between the thermal dependence of both physical properties allows us to conclude that deviation from C-W law at temperatures below *150K* is due to the thermal increase of $\xi(T)$, and that localised moment is the responsible of the magnetic behaviour of bulk Pd.

The thermal excitation rate of localised electrons that governs the increase of moment illustrated in Figure 3 should be determined by the partial density of states corresponding to the narrower or localised bands [23]. The total band structure of Pd, shown in Figure 4, is very well known since long time ago and has been determined theoretically [24-27] as well as experimentally by magneto-optical spectroscopy [19] and De Haas- van Alphen effect [28]. Just for an interval of approximately *0,5 eV* above the Fermi level the density of states, $g(\varepsilon)$, undergoes a drastic decrease of one order of magnitude from *1,23* to *0,25 states · eV$^{-1}$ · spin$^{-1}$ · atom$^{-1}$*. In the detailed band structure depicted in ref. [27] it can be seen a very narrow *d* sub-band waving around the Fermi level and, consequently, being un-filled and corresponding to localised states. Thermal transition from this localised band to the *s* band accounts for the magnetic moment growth.

Let us consider a particle with $N$ atoms and $N^*$ of them carrying uncorrelated and frozen permanent magnetic moment, $m$, randomly distributed in orientation. The spontaneous average moment per atom in the NP, $\langle m \rangle$, is then given by $\langle m \rangle = (N^*)^{1/2} \cdot \frac{m}{N}$. Note that for a ferromagnetic configuration of the $N^*$ moments the corresponding $\langle m \rangle$ should be $N^* \cdot \frac{m}{N}$; this indicates that for small $N^*$ the average moment for a randomly oriented system is of the same order to that corresponding to the ferromagnetic configuration. According to the values of $10^{-3} \mu_B \cdot at^{-1}$ reported in [1] for $\langle m \rangle$ and after considering $m = 1,5 \mu_B$ and $N = 10^3$ it is inferred, from (3), that $N^*$ is of the order of few units of atoms per NP. Twin boundaries at surfaces of NPs, that are shown to be present in a high amount for Pd [1], are sources of broken symmetry that can give rise to large local anisotropies. Large local anisotropy has been shown to be reinforced by spin-orbit coupling that induces orbital momentum, $L$, of the surface itinerant electrons around the localised spin [8]. The effective anisotropy of the localised moment associated with the spin is enhanced, according to [8], up to $\alpha \cdot L \cdot \hbar^2$, where the spin orbit coupling constant $\alpha \cdot \hbar^2$ is of the order of *0,4 eV* for Au and *0,05 eV* for Pd. Therefore, for *L = 1* the effective anisotropy can reach a giant value of *0,05 eV* per atom that will keep blocked the moment direction up to temperatures above 300 K, consequently explaining the constancy of the magnetization between 5 and 300 K, reported in [1]. The effective anisotropy field given by $\alpha \cdot \hbar^2 / \mu_B$ becomes then of the order of $10^2 T$.

Then, in contrast with the case of gold NPs that requires suitable capping to induce atomic magnetic moments [4], the existence of localised atomic moments in Pd leads to a permanent magnetism without any capping requirement, as recently described in detail [3].

In summary, it has been shown that i) Magnetism of bulk Pd is the sum of two contributions, a localised and an itinerant paramagnetism. Only localised Curie contribution is experimentally detected. ii) The thermal dependence of the localised moment evidences that some *d* band orbitals of bulk Pd must be considered as localised states and iii) The existence of permanent magnetism in Pd NPs can be envisaged as a consequence of the presence at their surfaces of localised atomic moments that are frozen in direction through its spin-orbit coupling to the itinerant electrons. This argument can be extended to any type of NPs with conduction electrons, high spin-orbit coupling and localised magnetic moments.

**Acknowledgments**

The authors are indebted to Prof. F. Yndurain for fruitful discussion and encouragement.

**Figure Captions**

**Figure 1.** Magnetic susceptibility of Pd as a function of the temperature measured at different fields.

**Figure 2.** (black) Inverse of the susceptibility of Pd vs T (measured at 1000 Oe) and (red) the fit to C-W law at high T, corresponding to the parameters C=0.422 K y Θ=-260 K.

**Figure 3.** Thermal evolution of the number of carriers for Pd atom obtained from Hall effect measurements (from ref. [14]) and the magnetic moment for Pd atom obtained form susceptibility measurements assuming a C-W law valid for all the range 5-300 K.

**Figure 4.** Density of states of Pd from ref. [19].

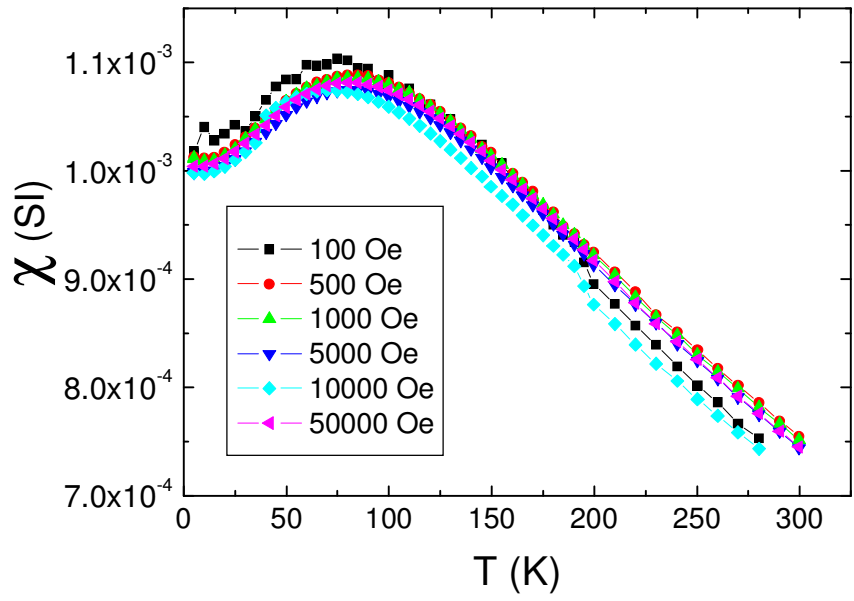

Figure 1 Hernando et al

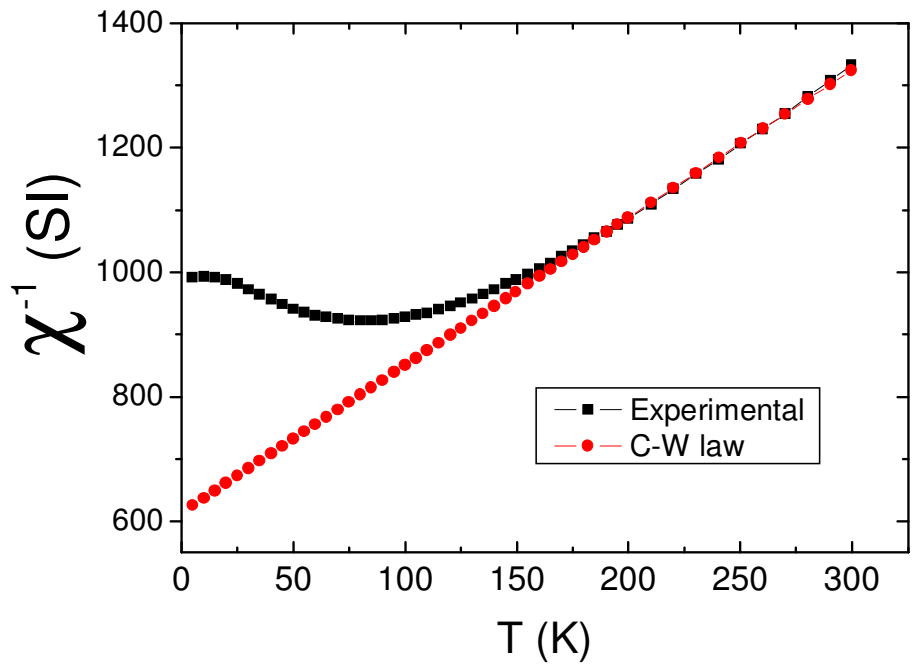

Figure 2 Hernando et al

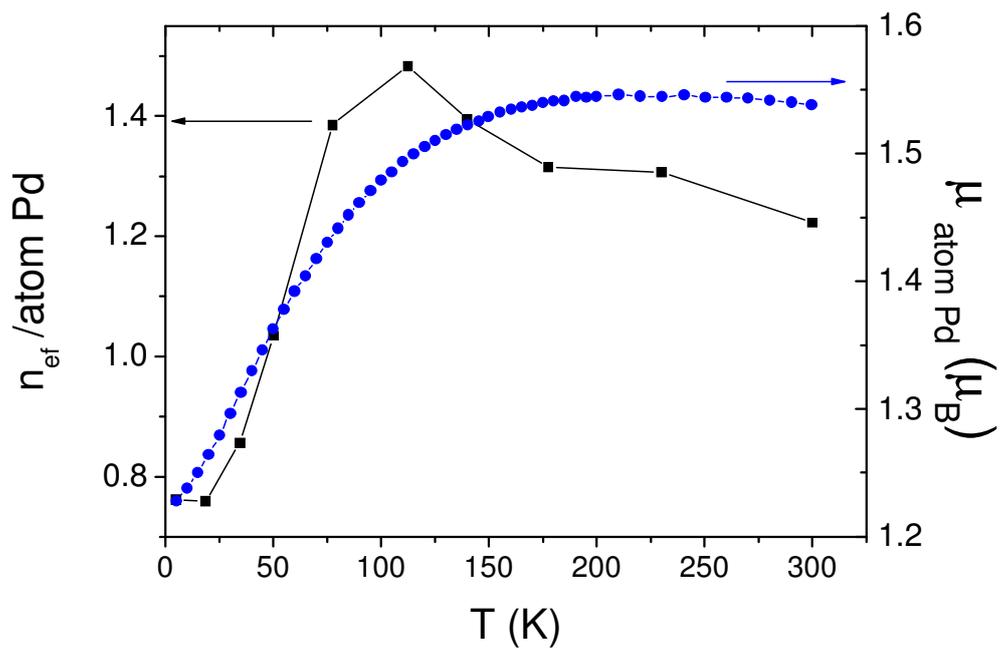

Figure 3 Hernando et al

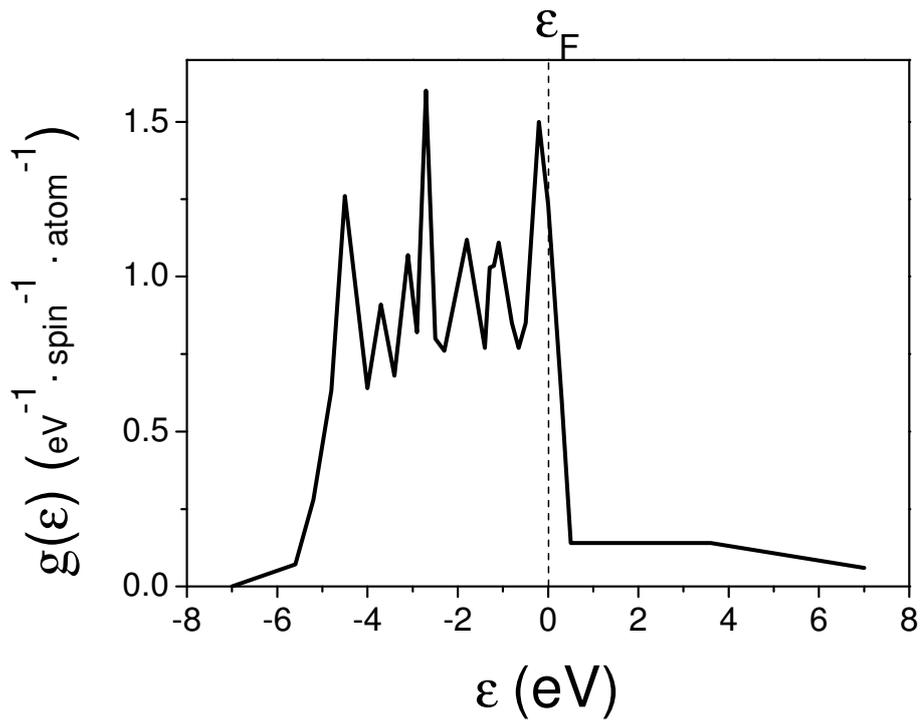

Figure 4 Hernando et al.